\documentclass[10pt]{article}
\usepackage[dvips]{color}
\usepackage{epsfig}
\usepackage{amsmath}
\usepackage{graphicx}

\begin{document}
\setcounter{page}{1}

\pagestyle{plain} \vspace{1cm}
\begin{center}
{\Large \bf Non-minimally Coupled Tachyon Field in Teleparallel
Gravity}\\ \small \vspace{1cm} {\bf Behnaz Fazlpour $^{a}$
\footnote{b.fazlpour@umz.ac.ir}} and {\bf Ali Banijamali $^{b}$
 \footnote{a.banijamali@nit.ac.ir}} \\
\vspace{0.5cm}   $^{a}$ {\it Department of Physics, Babol Branch,
Islamic Azad University, Babol, Iran\\} \vspace{0.5cm} $^{b}$ {\it
Department of Basic Sciences, Babol University of Technology, Babol,
Iran\\}
\end{center}
\vspace{1.5cm}
\begin{abstract}
We perform a full investigation on dynamics of a new dark energy
model in which the four-derivative of a non-canonical scalar field
(tachyon) is non-minimally coupled to the vector torsion. Our
analysis is done in the framework of teleparallel equivalent of
general relativity which is based on torsion instead of curvature.
We show that in our model there exists a late-time scaling attractor
(point $P_{4}$), corresponding to an accelerating universe with the
property that dark energy and dark matter densities are of the same
order. Such a point can help to alleviate the cosmological
coincidence problem. Existence of this point is the most significant
difference between our model and another model in which a canonical
scalar field (quintessence) is used instead of tachyon
field.\\


\end{abstract}

\newpage
\section{Introduction}
A mysterious component of our universe with negative pressure can be
a reasonable description for one of the most important discoveries
in cosmology. This component is known as dark energy (for reviews on
dark energy see [1, 2] and references therein) and the discovery is
the universe late-times acceleration supported by the cosmological
observations [3].\\
Besides the cosmological constant that suffers from cosmological
constant problem [4], the simplest candidate of dark energy is a
canonical scalar field, the so-called quintessence [5]. One can also
generalize quintessence model by including a non-minimal coupling
between scalar field and gravity [6-8]. Non-canonical scalar field
models of dark energy [9-13] and specially tachyon field model [14,
15]
also exist in the literature notably.\\
Furthermore, teleparallel gravity originally proposed by Einstein
[16, 17] is a theory of gravity which is based on torsion instead of
curvature formulation. In teleparallel gravity the curvature-less
Weitzenbock connection is used rather than the torsion-less
Levi-Civita one [16-19] and dynamical objects are the four linearly
independent vierbein (tetrad) fields. The teleparallel Lagrangian is
described by torsion scalar $T$. In Friedmann-Robertson-Walker (FRW)
space-time it is completely equivalent to a matter dominated
universe in the framework of general relativity. Thus, teleparallel
gravity without modification can not describe the accelerated
expansion of the universe. Inspired by the similar procedures in
general relativity there are two ways for such modification. The
first one is the construction of $f(T)$ gravity by extending $T$ to
an arbitrary function [20-23] and the second one is to directly add
dark energy into teleparallel gravity allowing also a non-minimal
coupling between dark energy and gravity. Note, however that in
$f(T)$ theories the action and field equations are not invariant
under local Lorentz transformations. This problem along with the
appearance of extra degrees of freedom with respect to general
relativity are crucial issues of $f(T)$ gravity [23]. Recently a
non-minimally coupled quintessence field in the framework of
teleparallel gravity, the so-called teleparallel dark energy model,
has been considered [24-28] and its generalization to a
non-canonical scalar field (tachyon) model has been studied in [29,
30]. In these models scalar field non-minimally coupled to torsion
scalar. As it was emphasized in [24] replacing the Ricci scalar $R$
by torsion scalar $T$ in the case of the minimal coupling of scalar
field dark energy models has no new result in the level of field
equations and perturbation as well. However, things are different if
we switch on the non-minimal coupling. In this case the resulting
coupled equations do not coincide. Clearly, teleparallel gravity
under the non-minimal coupling is a different theory.\\
In our previous works [46, 47] we have investigated dynamical
analysis of non-interacting and interacting tachyonic teleparallel
dark energy in details. In [46] we have extracted critical points of
the non-interacting scenario where they are all dark energy
dominated solutions. When interaction between dark energy and dark
matter was considered in the model, although late-time accelerated
attractor solutions have been found, there was no scaling attractor
[47].\\
Very recently inspired by the fact that in teleparallel gravity a
vector field interacts with the vector part of the torsion [31, 32],
Otalora has proposed a new dark energy model in which the
four-derivative (which is a vector field) of a canonical scalar
field (quintessence) couples non-minimally to the vector part of
torsion [33]. The author has studied dynamics of the model and its
cosmological implications in details. Here we generalize such a
model using tachyon field as a
responsible for dark energy.\\
The plan of the work is the following: In section 2 we present the
model and derive basic equations and expressions including energy
density, pressure and equation of motion of the scalar field. In
section 3 dynamical system study of the model is done and the
cosmological implications are discussed. Section 4 is devoted to our
conclusions.\\
\section{Tachyon Field in Teleparallel Gravity: Basic Equations}
Three components of torsion tensor $T_{\,\,\mu\nu}^{\lambda}$,
namely vector torsion, axial torsion and pure tensor are given as
follows respectively [33]
\begin{equation}
\mathcal{V}_{\mu}=T_{\,\,\nu\mu}^{\nu},
\end{equation}

\begin{equation}
\mathcal{A}^{\mu}=\frac{1}{6}\epsilon^{\mu\nu\rho\sigma}T_{\nu\rho\sigma},
\end{equation}

\begin{equation}
\mathcal{T}_{\lambda\mu\nu}=\frac{1}{2}\big(T_{\lambda\mu\nu}+T_{\mu\lambda\nu}\big)+\frac{1}{6}\big(g_{\nu\lambda}\mathcal{V}_{\mu}+
g_{\nu\mu}\mathcal{V}_{\lambda}\big)-\frac{1}{3}g_{\lambda\mu}\mathcal{V}_{\nu}.
\end{equation}
Our model with a non-minimal coupling between four-derivative of
tachyon field and vector torsion (1) can be described by the
following action
\begin{equation}
 S=\int d^{4}x\,h\,\left[\frac{T}{2\,\kappa^2}-V(\varphi)\,\sqrt{1-\partial_{\mu}\varphi\,\partial^{\mu}\varphi}+\eta\,f(\varphi)
 \partial_{\mu}\varphi\, \mathcal{V}^{\mu}\right]+S_{m},
\end{equation}
where $h\equiv\det(h^{a}_{~\mu})=\sqrt{-g}$ ( $h^{a}_{~\mu}$ are the
orthonormal tetrad components), $T$ is the torsion scalar
corresponding to teleparallel equivalent of general relativity
(TEGR) and $\kappa^{2}=8\pi G$ while $G$ is a bare gravitational
constant (for reviews on teleparallelism see [31, 32]). The second
part in the action (4) is the relevant Lagrangian of the tachyon
field and $S_{m}(\psi_{m},h^{a}\vspace{0.1 mm}_{\rho})$ is the
action of the matter field which is chosen as the cold dark matter
(DM). Also $\eta$ is a dimensionless constant measuring the
non-minimal coupling and $f(\varphi)$ is an arbitrary function of
the scalar field. In the other hand, note that there is no
interaction between dark matter and dark energy.\\
We mention that the model (4) is a generalization of the model
proposed in [33] by using a non-canonical scalar field (tachyon)
instead of canonical scalar field (quintessence). Tachyon field
originates from a fundamental theory such as string theory and has
interesting cosmological implications (see for example [14, 15]).\\
In the context of general relativity, non-minimal couplings are
generated by quantum corrections to the scalar field theory and they
are essential for the renormalizability of the scalar field theory
in curved space (see [39] and references therein). In the other
side, non-minimal coupling between derivatives of a scalar field and
curvature may appear in some Kaluza-Klein theories [40, 41].\\
The gravitational coupling of the fundamental fields in teleparallel
gravity is a very controversial subject. Since there is no
experimental data to help us, we should rely on equivalence between
general relativity and teleparallel gravity. According to this
formulation, each one of the fundamental fields of nature (scalar,
spinor, and electromagnetic) are required to couple to torsion in a
such a way to preserve the equivalence between teleparallel gravity
and general relativity [31]. It is shown in [32] that in the context
of teleparallel gravity a scalar field by itself does not feel
gravity but its four-derivative (which is a vector field) interacts
with the vector part of
torsion.\\
As it is noted in [33] the model in (4) and tachyonic teleparallel
dark energy model [29] in which tachyon field non-minimally coupled
to torsion scalar instead of vector torsion, are mathematically
related through a conformal transformation although they are
physically different.\\
In the other side, the possibility that the dynamic Poincare gauge
theory connection, reflected in dynamic Poincare gauge theory
torsion, provides the accelerating force in the universe has been
explored by Shie et al. in [38]. The scalar mode torsion in the
model plays the role of the imperceptible dark energy and it is
naturally obtained from the geometry of the Riemann-Cartan
space-time, instead of from an exotic scalar field or a designed
mechanism. Note however, as it is mentioned in [38] if we consider
the space-time as Riemannian instead of Riemann-Cartan, by absorbing
the contribution of the torsion of this model into the stress-energy
tensor on the rhs of the Einstein equation, then this contribution
will act as a source of the Riemannian metric, effectively like an
exotic fluid with negative pressure which drives the universe into
accelerating expansion. But in the present paper, dark energy is
attributed to a scalar field (tachyon) in the framework of
teleparallel gravity in which one replaces the Riemannian curvature
scalar $R$ by the scalar torsion of teleparallel gravity allowing
also a non-minimal coupling between the four-derivative of the
scalar field with vector part of
torsion tensor.\\
To study tachyon dynamics more simply let us apply the following
transformation (field redefinition) in action (4) [34]
\begin{equation}
\varphi \rightarrow \phi=\int d\varphi \sqrt{V(\varphi)}
\Longleftrightarrow
\partial \varphi=\frac{\partial \phi}{\sqrt{V(\phi)}}.
\end{equation}
It leads to our starting action as follows
\begin{equation}
 S=\int d^{4}x\,h\,\left[\frac{T}{2\,\kappa^2}-V(\phi)\,\sqrt{1-\frac{\partial_{\mu}\phi\,\partial^{\mu}\phi}{V(\phi)}}+\frac{\eta\,f(\phi)
 \partial_{\mu}\phi\, \mathcal{V}^{\mu}}{\sqrt{V(\phi)}}\right]+S_{m}.
\end{equation}
Considering a spatially-flat FRW metric,
\begin{eqnarray}
ds^{2}=dt^{2}-a^{2}(t)(dr^{2}+r^{2}d\Omega^{2}),
\end{eqnarray}
and a tetrad choice of the form $h^{a}_{\mu}=diag(1,a,a,a)$ (which
is a solution of gravitational field equation [17]), and a
homogeneous scalar field $\phi$ yield to the Friedmann and
Raychaudhuri equations as follows,
\begin{equation}
H^{2}=\frac{\kappa^{2}}{3}\big(\rho_{\phi}+\rho_{m}\big),
\end{equation}
\begin{equation}
\dot{H}=-\frac{\kappa^{2}}{2}\big(\rho_{\phi}+P_{\phi}+\rho_{m}+P_{m}\big),
\end{equation}
where $H=\frac{\dot{a}}{a}$ is the Hubble parameter, $a$ is the
scale factor and a dot stands for derivative with respect to the
cosmic time $t$. $\rho_{m}$ and $P_{m}$ are the matter energy
density and pressure respectively, satisfying the equation
$\dot{\rho}_{m}+3H(1+\omega_{m})\rho_{m}=0$, with
$\omega_{m}=\frac{P_{m}}{\rho_{m}}$ the matter equation of state
parameter.\\
The energy momentum tensor associated with the scalar field
$\Theta_{a}\vspace{0.1 mm}^{\rho}\equiv-\frac{1}{h}\frac{\delta
S_{\phi}}{\delta h^{a}\vspace{0.1 mm}_{\rho}}$ is given by

$$\Theta_{a}\vspace{0.1
mm}^{\rho}=\eta\big[\frac{f(\phi)}{\sqrt{V}}(\mathcal{V}^{\rho}\partial_{a}\phi+\nabla_{a}\partial^{\rho}\phi
-h_{a}\vspace{0.1
mm}^{\rho}\nabla_{\mu}\partial^{\mu}\phi)+\big(\frac{f_{,\phi}}{\sqrt{V}}-\frac{f(\phi)V_{,\phi}}{2V^{\frac{3}{2}}}\big)
(\partial_{a}\phi\,\partial^{\rho}\phi-h_{a}\vspace{0.1
mm}^{\rho}\partial_{\mu}\phi\,\partial^{\mu}\phi)\big]$$
\begin{equation}
-\mu^{-1}V(\phi)\delta_{a}\vspace{0.1 mm}^{\rho}-\mu\,
\partial_{a}\phi\,\partial^{\rho}\phi,
\end{equation}
where $\nabla^{\mu}$ is the covariant derivative in the teleparallel
connection [31, 32], $f_{,\phi}= \frac{d f(\phi)}{d\phi}$,
$V_{,\phi}=\frac{dV}{d\phi}$ and
$\mu=\frac{1}{\sqrt{1-\frac{\dot{\phi}^{2}}{V}}}$.\\
By imposing the FRW metric (7), the energy density and pressure of
the scalar field read,
\begin{equation}
 \rho_{\phi}=\mu\,V\left(\phi\right)-3\,\eta\,\frac{f(\phi)}{\sqrt{V}}\,H\,\dot{\phi},
\end{equation}
and
\begin{equation}
 p_{\phi}=-\mu^{-1}\,V(\phi)+\eta\,\dot{\phi}^{2}\Big(\frac{f_{,\phi}}{{\sqrt{V}}}-\frac{f(\phi)V_{,\phi}}{2 V^{\frac{3}{2}}}\Big)
 +\eta
 \frac{f(\phi)}{\sqrt{V}}\,\ddot{\phi},
\end{equation}

Additionally, variation of the action (6) with respect to the scalar
field yields to its evolution equation that in FRW background takes
the form
\begin{equation}
 \ddot{\phi}\Big[\big(\frac{\partial\mathcal{L}}{\partial X}\big)+(2X)\big(\frac{\partial^{2}\mathcal{L}}{\partial X^{2}}\big)\Big]
 +\Big[3H\big(\frac{\partial\mathcal{L}}{\partial X}\big)+\dot{\phi}\big(\frac{\partial^{2}\mathcal{L}}{\partial X \partial\phi}\big)\Big]
 \dot{\phi}-\big(\frac{\partial\mathcal{L}}{\partial \phi}\big)=0,
\end{equation}
where $X=\frac{1}{2}\dot{\phi}^{2}$.\\
From equation (13), one can obtain the $\phi$-filed equation of
motion more clearly as follows,
\begin{equation}
 \ddot{\phi}+3\,\mu^{-2}\,H\,\dot{\phi}+\left(1- \frac{3\,\dot{\phi}^{2}}{2V}\right)V_{,\phi}-3\,\eta\,\left(\dot{H}+3H^{2}\right)\,
 \mu^{-3}\,\frac{f(\phi)}{\sqrt{V}}=0.
\end{equation}
In fact the above equation expresses the energy conservation
relation
$\dot{\rho}_{\phi}+3\,H\left(1+\omega_{\phi}\right)\rho_{\phi}=0$
with $\omega_{\phi}= p_{\phi}/\rho_{\phi}$ the equation of state of
the scalar field.

\section{Dynamical Analysis}
Now we are going to rewrite the evolution equations as a system of
autonomous differential equations and study the cosmological
properties of the critical points for the system. An autonomous
system in general will be of the form $\frac{d\textbf{Y}}{d \ln
a}=f(\textbf{Y})$, where the column vector $\textbf{Y}$ is
constituted by suitable auxiliary variables and $f(\textbf{Y})$ is
the corresponding column vector of the autonomous
equations [35-37].\\
The critical points $\textbf{Y}_{c}$ are obtained from
$\frac{d\textbf{Y}}{d \ln a}=0$. In order to study the stability of
the equilibrium or critical points we should first expand the system
around $\textbf{Y}_{c}$ as $\textbf{Y}=\textbf{Y}_{c}+\textbf{U}$
where the column vector $\textbf{U}$ denotes the perturbation of the
variables. For each critical point the 1st order perturbation
technique lends to the matrix equation $\textbf{U}'=\Sigma .
\textbf{U}$ where the matrix $\Sigma$ contains all the coefficients
of the perturbation equations. Hence, the stability properties and
the type of a specific critical point are determined by the
eigenvalues of the matrix $\Sigma$.\\
(i) If the real part of eigenvalues have opposite signs then the
fixed point is a saddle point.\\
(ii) If three eigenvalues are negative, then the fixed point is
stable.\\
(iii) If the eigenvalues are positive, the fixed point in
unstable.\\
A detailed analysis of the stability criteria is given in Refs
[35-37]. A critical point is an attractor in the case (ii) and the
universe evolves to the attractor solutions regardless of the
initial conditions.\\
Let us apply the above method to the model (6) by introducing the
following new variables:
\begin{equation}
 x\equiv\frac{\dot{\phi}}{\sqrt{V}}, \:\:\:\:\:\: y\equiv\frac{\kappa\,\sqrt{V}}{\sqrt{3}\,H}, \:\:\:\:\:\:
   u\equiv\frac{\kappa\,\sqrt{f}}{\sqrt{H}}, \:\:\:\:\ \alpha\equiv\frac{f_{,\phi}}{\kappa f}, \:\:\:\: \lambda\equiv-\frac{V_{,\phi}}{\kappa V}.
\end{equation}
In terms of these new variables, the field equations can be written
as follows,
\begin{equation}
 \frac{dx}{dN}=\left(1-x^{2}\right)\left[(3-s)\eta\,\mu^{-1}\,u^{2}y^{-2}+\sqrt{3}\left(\lambda\,y-\sqrt{3}\,x\right)\right],
  \end{equation}
\begin{equation}
 \frac{dy}{dN}=\left(-\frac{\sqrt{3}\,\lambda}{2}\,x\,y+s\right)\,y,
\end{equation}

\begin{equation}
 \frac{du}{dN}=\frac{1}{2}\left(\sqrt{3}\,\alpha\,x\,y+s\right)\,u,
\end{equation}

\begin{equation}
 \frac{d \lambda}{dN}=-\sqrt{3}\,\lambda^{2}\,x\,y\,\left(\Gamma-1\right),
 \end{equation}

\begin{equation}
 \frac{d \alpha}{dN}=\sqrt{3}\,\alpha^{2}\,x\,y\,\left(\Pi-1\right),
 \end{equation}
 where $N=\ln{a}$ and the following
 parameters are defined
\begin{equation}
\Pi=\frac{f\,f_{,\phi\phi}}{f_{,\phi}^{2}}, \:\:\:\:\: \:\:\:\:\:\:
\Gamma=\frac{V\,V_{,\phi\phi}}{V_{,\phi}^2}.
\end{equation}
 In equation (16)-(18), $s$ is given by
\begin{multline}
 s=-\frac{\dot{H}}{H^{2}}=3\,\left(2+\eta^{2}\mu^{-3}u^{4}y^{-2}\right)^{-1}\\
\left[\gamma-(\gamma-1)\left(
 \mu\,y^{2}-\eta\,u^{2}\,x\right)-\mu^{-1}\,y^{2}+\eta\,u^{2}\left(-\mu^{-2}\,x+\frac{\sqrt{3}}{3}y
 \left(x^{2}(\alpha-\lambda)+\lambda\right)+\eta\,\mu^{-3}u^{2}y^{-2}\right)\right],
\end{multline}
and we mention that $\gamma$ is the barotropic index defined by
$\gamma=1+\omega_{m}$ such that $0<\gamma<2$
and also $\mu=\frac{1}{\sqrt{1-x^{2}}}$.\\
The density parameters
$\Omega_{i}\equiv(\kappa^{2}\,\rho_{i})/(3\,H^{2})$ for the scalar
field and background matter are given by
\begin{equation}
 \Omega_{\phi}=\mu\,{y}^{2}-\eta\,x\,{u}^{2}, \:\:\:\:\:\:\:\:
 \Omega_{m}=1-\Omega_{\phi},
\end{equation}
while the equation of state of the field $\omega_{\phi}$ and the
effective equation of state can be written as
\begin{equation}
 \omega_{\phi}=\frac{p_{\phi}}{\rho_{\phi}}=\frac{-\,\mu^{-1}\,{y}^{2}+\eta\,u^{2}\left(-\mu^{-2}\,x+\frac{\sqrt{3}}{3}y
 \left(x^{2}(\alpha-\lambda)+\lambda\right)+\frac{1}{3}\eta\,\mu^{-3}\,(3-s)\,u^{2}y^{-2}\right)}
 {\mu\,{y}^{2}-\eta\,x\,{u}^{2}},
 \label{21}
\end{equation}
and
\begin{multline}
\omega_{eff}=\left(p_{\phi}+p_{m}\right)/\left(\rho_{\phi}+\rho_{m}\right)\\
=\left({x}^{2}-\gamma\right)\mu
\,{y}^{2}+\eta\,u^{2}\left(-\mu^{-2}\,x+\frac{\sqrt{3}}{3}y
\left(x^{2}(\alpha-\lambda)+\lambda\right)+\frac{1}{3}\eta\,\mu^{-3}
\,(3-s)\,u^{2}y^{-2}\right)+\,(\gamma-1)\left(1+\eta\,x\,u^{2}\right).
\end{multline}
Note that the condition for acceleration is $\omega_{eff}<-1/3$.\\
Once the parameters $\Gamma$ and $\Pi$ are known, equations
(16)-(20) become a system of autonomous differential equations and
one can study dynamics of the model in a usual way. Considering an
exponential potential of the form $V=V_{0} \,e^{-\lambda\kappa\phi}$
with constant $\lambda$ leads to $\Gamma=1$ and equation (19) can be
eliminated from our system of differential equations. In the other
hand if we consider a constant $\alpha$ for simplicity then the
number of differential equations will be reduced once again.
Therefore, we concentrate on a non-minimal coupling function of the
form $f(\phi)\propto \,e^{\beta\phi}$ with a constant $\beta$ which
leads to $\Pi=1$ and $\alpha$ becomes a constant. The exponential
form of the scalar field potential has been found in higher-order
[43] or higher-dimensional gravity theories [44]. In string or
Kaluza-Klein type models the moduli fields associated with the
geometry of the extra dimensions may have effective exponential
potentials due to curvature of the internal spaces, or the
interaction of moduli with form fields on the internal spaces.
Exponential potentials can also arise due to nonperturbative effects
such as gaugino condensation [45]. Significant role of an
exponential potential can be seen in various cosmological models
(for example see [35]). Moreover, if we consider other forms of $V$
and $f$ the number of autonomous equations will be increased.
Extracting the fixed points of a 4 or 5-dimensional autonomous
system is pretty complicated. So, by choosing the potential $V=V_{0}
\,e^{-\lambda\kappa\phi}$ and non-minimal coupling function
$f(\phi)\propto \,e^{\beta\phi}$, we have a three dimensional
autonomous system (16)-(18).\\
The critical points of the autonomous system (16)-(18) with constant
$\alpha$ and $\lambda$ are listed in Table 1. In the same table we
have provided the corresponding values of $\Omega_{\phi}$,
$\omega_{\phi}$ and $\omega_{eff}$ at each critical point. The
conditions needed for existence and acceleration
$\big(\omega_{eff}<-1/3\big)$ along with stability properties of the
fixed points presented in Table 2. The elements of $3\times 3$
matrix $\Sigma$ and its three eigenvalues at each critical point are
shown in Appendix. Now let us discuss the cosmological behavior and
stability properties of the critical
points individually.\\
\\
\textbf{Critical Point $P_{1}$:}\\
This point exists for positive values of $\lambda$ and accelerated
expansion of our universe occurs for $\gamma<\frac{2}{3}$. Point
$P_{1}$ is stable for $\alpha<-\frac{\lambda}{2}$ and
$\lambda>\sqrt{\frac{3\gamma}{\sqrt{1-\gamma}}}$ and thus attracts
the universe at the late-times. Since the dark energy and matter
density parameters are at the same order, this point corresponds to
a dark energy-dark matter scaling solution, alleviating the
cosmological coincidence problem which asks: why are we living in an
epoch in which $\Omega_{DE}$ and $\Omega_{DM}$ are comparable? In
order to show the above mentioned behavior more transparently, we
evolve the autonomous system (16)-(18) numerically for the parameter
choices $\lambda=2$ and $\alpha=-1.5$. The phase-space trajectories
are depicted in a 3-dimensional plot in Figure 1 (left panel). It is
clear from the figure that in this case the universe is attracted by
the stable solution $P_{1}$. An important point about $P_{1}$ is
that, it is not a realistic solution in dark energy paradigm due to
the presence of the condition $\gamma<\frac{2}{3}$ needed for
acceleration. One can solve such a problem by considering a possible
coupling between dark energy and dark matter. We will consider this
possibility in future works.\\
\textbf{Critical Point $P_{2}$:}\\
This point is a stable point for $\alpha>-\frac{\lambda}{2}$ and
$\lambda<-\sqrt{\frac{3\gamma}{\sqrt{1-\gamma}}}$ and therefore it
can be the late-times state of the universe. Similar to point
$P_{1}$, $P_{2}$ can be accelerated for $\gamma<\frac{2}{3}$ and
since $\Omega_{\phi}$ and $\Omega_{m}$ are of the same order it is a
scaling attractor (it can help to alleviate the cosmological
coincidence problem).\\
Depending on the background matter (i.e. the values of $\gamma$) the
dark energy equation of state lies in the quintessence regime
$(\omega_{\phi}>-1)$ or is equal to cosmological constant value
$(\omega_{\phi}=-1)$. In Figure 1 (right panel) the phase-space
trajectories of the system (16)-(18) for the model parameter values
$\lambda=-2$, $\alpha=1.5$ and $\eta=\frac{1}{2}$ are presented. In
this case the attractor point of the system is $P_{2}$ as it is
clear from the figure. Once again note that $P_{2}$ is also a
non-realistic solution because of the condition
$\gamma<\frac{2}{3}$ which is needed for acceleration.\\
\textbf{Critical Point $P_{3}$:}\\
The fixed point $P_{3}$ corresponds to a dark energy dominated de
Sitter solution with $\Omega_{\phi}=1$ and $\omega_{eff}=-1$. This
point exists when $\frac{\lambda}{\eta}<0$ and can be accelerated
for all values. It is an attractor solution of the model for
$\lambda>0$, $\alpha>-\frac{\lambda}{2}$ or $\lambda<0$,
$\alpha<-\frac{\lambda}{2}$. Such a behavior of the system (16)-(18)
are depicted in Figure 2 for the choices $\lambda=2$, $\alpha=1$ and
$\eta=-\frac{1}{2}$ (left panel) and $\lambda=-2$,
$\alpha=0.5$ and $\eta=\frac{1}{2}$ (right panel) respectively.\\
\textbf{Critical Point $P_{4}$:}\\
At this point $\Omega_{\phi}$ and $\Omega_{m}$ are of the same order
and thus it can help to alleviate the cosmological coincidence
problem. $P_{4}$ exists for positive values of $\lambda$ and
universe at this point can be accelerated for the conditions
presented in Table 2. However, since the eigenvalues of the $3\times
3$ matrix $\Sigma$ of the corresponding linearized perturbation
equations at this point are very complicated, we can not conclude
about its stability analytically. However we evolved the system
numerically and find that as long as $\alpha<-\frac{\lambda}{2}$
this point can be a stable point as it is shown in Figure 3. In this
figure the parameters of the model have the values $\lambda=3$,
$\alpha=-2$ and $\eta=\frac{1}{2}$.\\
\textbf{Critical Point $P_{5}$:}\\
This critical point has the same acceleration and eigenvalues
properties as point $P_{4}$ but it exists for negative values of
$\lambda$. Our numerical computations reveal that $P_{5}$ could not
be a stable solution of the model.\\
The cosmological evolutions of the density fraction parameters
$\Omega_{\phi}$ and $\Omega_{m}$ are also shown in Figure 4.\\
Before closing this section, let us make comments on the results of
the past models with a non-minimally coupled scalar field in the
framework of teleparallel gravity. Dynamics of teleparallel dark
energy in which a canonical scalar field (quintessence)
non-minimally coupled to torsion scalar was first studied by Wei in
[27]. The Author has considered the interaction between dark energy
and dark matter and for two kinds of potentials has obtained some
attractor points. But no scaling attractor was found. Xu et al. [26]
have investigated phase space analysis of the model using a new set
of dimensionless variables. Although they have found a dark energy
dominated solution with $\omega=-1$, unfortunately at this point
$\Omega_{\phi}=1$ and so it
is not a scaling attractor of the model.\\
Finally a scaling attractor of teleparallel dark energy model has
been found in Ref. [28]. The coupling function between scalar field
and torsion scalar was generalized to a arbitrary function $f(\phi)$
and interaction between dark energy and dark matter was also
considered.\\
Furthermore, dynamics of tachyonic teleparallel dark energy has been
studied in [30] and although scaling solutions in this model have
been found, they were not stable solutions. In the other hand stable
solutions in tachyonic teleparallel dark energy are not scaling
attractors of the model. Phase-space analysis of a new teleparallel
dark energy model has been done in [33]. In such a model instead of
tachyon, quintessence plays the role of dark energy in action (4). A
de Sitter attractor solution was found in the case that the
non-minimal coupling function is of the form $f(\phi)\propto\phi$
and a attractor solution was found for general form of $f(\phi)$.
But no scaling attractor was presented in the model. In the present
paper we have utilized a non-canonical scalar field (tachyon) in
action (4) and extracted the fixed points of the model. In addition
to dark energy dominated solution ($P_{3}$) that accelerates the
universe, we have obtained a scaling attractor solution ($P_{4}$)
that fulfills conditions required for
current state of our universe.\\

\begin{table}[t]
\caption{The critical points of the autonomous system (16)-(18) for
constant $\alpha$ and the corresponding values of the dark energy
density parameter $\Omega_{\phi}$, the dark energy equation of state
parameter $\omega_{\phi}$ and the effective equation of state
parameter $\omega_{eff}$. We use the notation
$A=-\lambda^{2}+\sqrt{\lambda^{4}+36}$.}
 \centering
\begin{center}
\begin{tabular}{|c|c|c|c|c|c|c|}\hline
Name & $x_{c}$ & $y_{c}$ & $u_{c}$  & $\Omega_{\phi}$ &
$\omega_{\phi}$ &$\omega_{eff}$\\
\hline $P_{1}$ & $\sqrt{\gamma}$ & $\frac{\sqrt{3\gamma}}{\lambda}$
& $0$ & $\frac{3\gamma}{\lambda^{2}\sqrt{1-\gamma}}$ & $\gamma-1$ &
$\gamma-1$\\\hline
 $P_{2}$ & $-\sqrt{\gamma}$ &
$-\frac{\sqrt{3\gamma}}{\lambda}$ & $0$ &
$\frac{3\gamma}{\lambda^{2}\sqrt{1-\gamma}}$ & $\gamma-1$ &
$\gamma-1$\\\hline $P_{3}$ & $0$ & $1$ &
$\sqrt{\frac{-\sqrt{3}\lambda}{3\eta}}$ & $1$ & $-1$ & $-1$\\\hline
$P_{4}$ & $\frac{\sqrt{2}}{6}\lambda \sqrt{A}$ &
$\frac{\sqrt{6}}{6}\sqrt{A}$ & $0$ &
$\frac{1}{6}\frac{A}{\sqrt{1-\frac{\lambda^{2}}{18}A}}$ &
$\frac{\lambda^{2}}{18}A-1$ &
$\frac{1}{6}\frac{A\left(\frac{\lambda^{2}}{18}A-\gamma\right)}{\sqrt{1-\frac{\lambda^{2}}{18}A}}+\gamma-1$\\\hline
$P_{5}$ & $-\frac{\sqrt{2}}{6}\lambda\sqrt{A}$ &
$\frac{\sqrt{6}}{6}\sqrt{A}$ & $0$ &
$\frac{1}{6}\frac{A}{\sqrt{1-\frac{\lambda^{2}}{18}A}}$ &
$\frac{\lambda^{2}}{18}A-1$ &
$\frac{1}{6}\frac{A\left(\frac{\lambda^{2}}{18}A-\gamma\right)}{\sqrt{1-\frac{\lambda^{2}}{18}A}}+\gamma-1$\\\hline
\end{tabular}
\end{center}
\end{table}

 \begin{table}
  \caption{Existence, acceleration and stability conditions of the fixed points in Table 1.}
 \centering
\begin{center}

\begin{tabular}{|c|c|c|c|}

   \hline label & existence & acceleration & stability  \\
  \hline $P_{1}$& $\lambda>0$&$\gamma<\frac{2}{3}$&$\begin{array}{c}
 stable\,\,point\,\, if \\
                    \alpha<-\frac{\lambda}{2}\,\,\,
                       and\,\,\,
                       \lambda>\sqrt{\frac{3\gamma}{\sqrt{1-\gamma}}} \\
                        \hline\\
                         saddle\,\,
                         point\,\,if\\
                         \alpha>-\frac{\lambda}{2}\,\,\,
                                             or\,\,\,
                         \lambda<\sqrt{\frac{3\gamma}{\sqrt{1-\gamma}}} \\
                           \end{array}$\\
 \hline $P_{2}$&$\lambda<0$&$\gamma<\frac{2}{3}$&$\begin{array}{c}
 stable\,\,point\,\, if \\
                    \alpha>-\frac{\lambda}{2}\,\,\,
                       and\,\,\,
                       \lambda<-\sqrt{\frac{3\gamma}{\sqrt{1-\gamma}}} \\
                        \hline\\
                         saddle\,\,
                         point\,\,if\\
                         \alpha<-\frac{\lambda}{2}\,\,\,
                                             or\,\,\,
                         \lambda>-\sqrt{\frac{3\gamma}{\sqrt{1-\gamma}}} \\
                           \end{array}$\\
     \hline $P_{3}$& $\frac{\lambda}{\eta}<0$& all values&$\begin{array}{c}
 stable\,\,point\,\, if \\
                     \lambda>0\,\,\,and\,\,\,\alpha>-\frac{\lambda}{2} \\
                        or\\
                       \lambda<0\,\,\,and\,\,\,\alpha<-\frac{\lambda}{2} \\
                        \hline\\
                         saddle\,\,
                         point\,\,if\\ \lambda<0\,\,\,and\,\,\,\alpha>-\frac{\lambda}{2}
                                             \\
                                             or\\
                         \lambda>0\,\,\,and\,\,\,\alpha<-\frac{\lambda}{2} \\
                           \end{array}$\\
         \hline $P_{4}$ &$\lambda>0$&$\gamma<\frac{2\left(2\sqrt{1-\frac{\lambda^{2}}{18}A}-\frac{\lambda^{2}A^{2}}{36}\right)}{6\sqrt{1-\frac{\lambda^{2}}{18}A}
         -A}$&see explanations about this point in the text\\

         \hline $P_{5}$ &$\lambda<0$&$\gamma<\frac{2\left(2\sqrt{1-\frac{\lambda^{2}}{18}A}-\frac{\lambda^{2}A^{2}}{36}\right)}
         {6\sqrt{1-\frac{\lambda^{2}}{18}A}-A}$&see explanations about this point in the text\\

 \hline

\end{tabular}
\end{center}
\end{table}
\section{Conclusion}
According to observational data we are living in an epoch in which
the densities of dark energy and dark matter are comparable [4],
although they scale differently during the expansion history of our
universe. This cosmological coincidence problem can be alleviated in
most dark energy models via the method of scaling attractors that
correspond to accelerating universe and ratio
$\frac{\Omega_{\phi}}{\Omega_{m}}$ of order $1$. If these conditions
are fulfilled, then the universe will result to that solution at
late-times, independently of the initial conditions and the
basic observational requirement will be satisfied [42].\\
In the present paper, motivated by the recent work of Otalora [33],
we proposed a new model of dark energy in which the four-derivative
of tachyon field is non-minimally coupled to the vector part of
torsion tensor. As it is mentioned in [33], such a non-minimal
coupling has no analogue in general relativity, because in general
relativity decomposition of gravitational field in a form analogous
to the decomposition of torsion in teleparallel gravity, is not
possible.\\
We studied the dynamical behavior of our model in details. In order
to reduce the number of autonomous equations, we have chosen an
exponential tachyonic potential $V=V_{0}e^{-\lambda\kappa\phi}$ and
a non-minimal coupling function of the form $f(\phi)\propto
e^{\beta\phi}$. These choices leave a three dimensional autonomous
equations out of the field equations (8), (9) and (14). We found
five critical points presented in Table 1. Points $P_{1}$ and
$P_{2}$ are scaling attractors of the model for suitable choices of
the model parameters as shown in Table 2 and depicted in Figure 1.
However these points have a disadvantage that acceleration occurs
for $\gamma<\frac{2}{3}$ which makes them non-realistic points in
applying to dark energy.\\
Point $P_{3}$ corresponds to a completely dark energy dominated
solution that can be accelerated but it is not a scaling attractor
of the model because at this point $\Omega_{\phi}=1$. The equation
of state at this point behaves like a cosmological constant
regardless of the values of the model parameters. The most
interesting critical point is the point $P_{4}$ which is a scaling
attractor that can be accelerated for the conditions presented in
Table 2. Thus point $P_{4}$ can help to alleviate the cosmological
coincidence problem and fulfills requirement according to
cosmological observations. Here we mention that in Ref [33] where
quintessence scalar field plays the role of dark energy, there is no
scaling attractor that can be accelerated. In comparison to
tachyonic teleparallel dark energy [46, 47], once again we emphasize
that in our previous works tachyon field non-minimally coupled to
the torsion scalar but here the field non-minimally coupled to the
vector torsion. The significant advantage of the present model is
that while in the previous models [46, 47] there was no scaling
attractor here we obtain a scaling attractor without considering an
extra term such as interaction between dark energy and dark matter
as usually done in the literature to obtain scaling attractors.
Therefore, Point $P_{4}$ makes an important difference between our
model and the models proposed in [33, 46, 47]. And finally point
$P_{5}$ is not a stable solution of the system (16)-(18). Before
closing this section we mention that besides the model (6),
generalized tachyon field in the context of teleparallel gravity
also represents scaling attractors [48]. Thus, in order to obtain a
scaling attractor through tachyonic teleparallel dark energy model
we should either generalized tachyon action [48] or consider
non-minimal coupling of tachyon field with vector torsion. Further
study of our model can be done for other scalar field potentials
$V(\phi)$ and coupling function $f(\phi)$.\\

\newpage
\begin{figure}[htp]
\begin{center}
\includegraphics{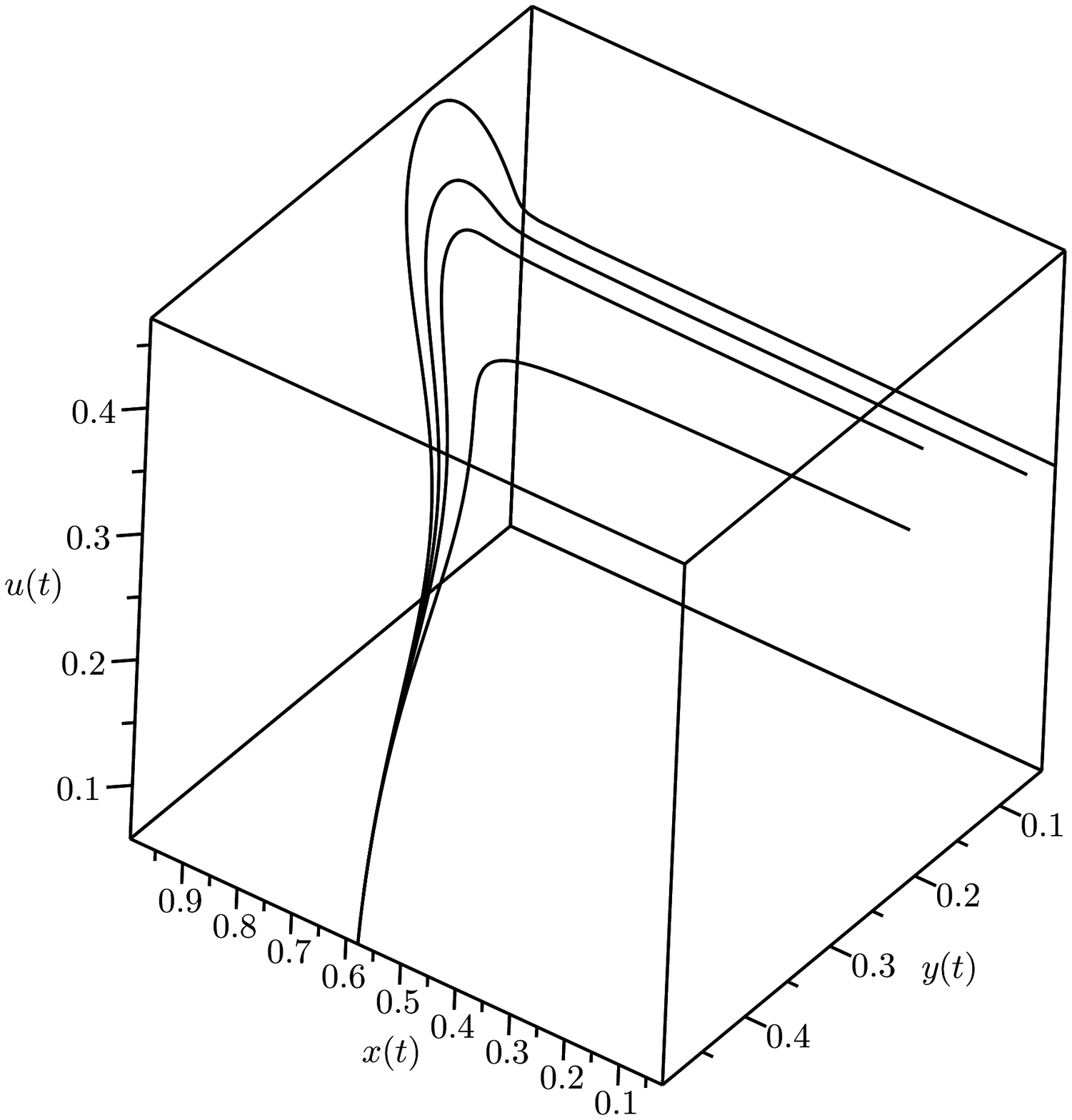} \vspace{2.5cm}\includegraphics{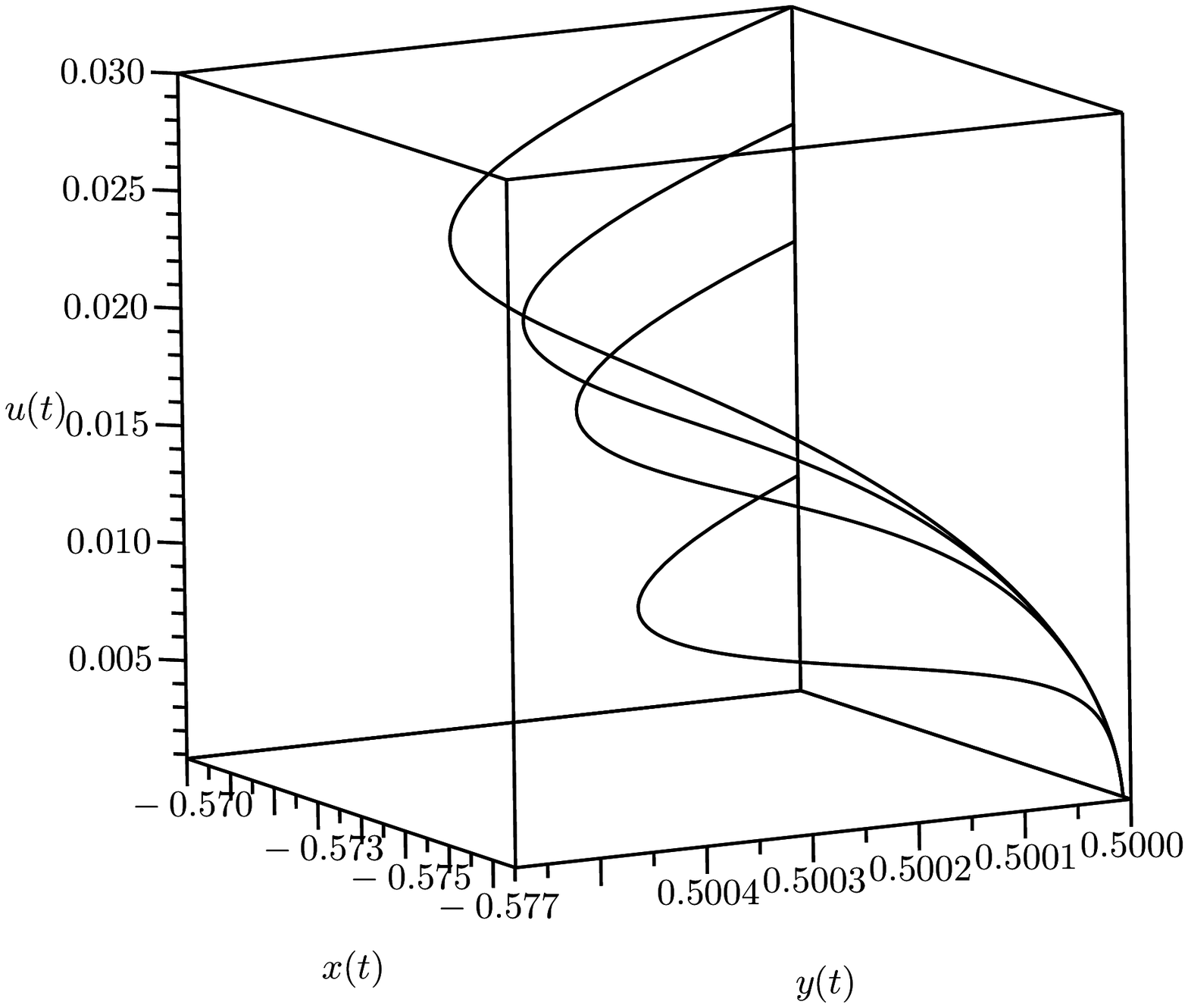}\vspace{2.5cm}

\caption{\small {3-dimensional phase-space trajectories of the
cosmological scenario (16)-(18) with stable attractor $P_{1}$ (left)
for the parameter choices $\lambda=2$, $\alpha=-1.5$ and
$\eta=\frac{1}{2}$ and stable attractor $P_{2}$ (right) for
$\lambda=-2$, $\alpha=1.5$ and $\eta=\frac{1}{2}$. }}
\end{center}
\end{figure}
\vspace{3cm}
\begin{figure}[htp]
\begin{center}
\includegraphics{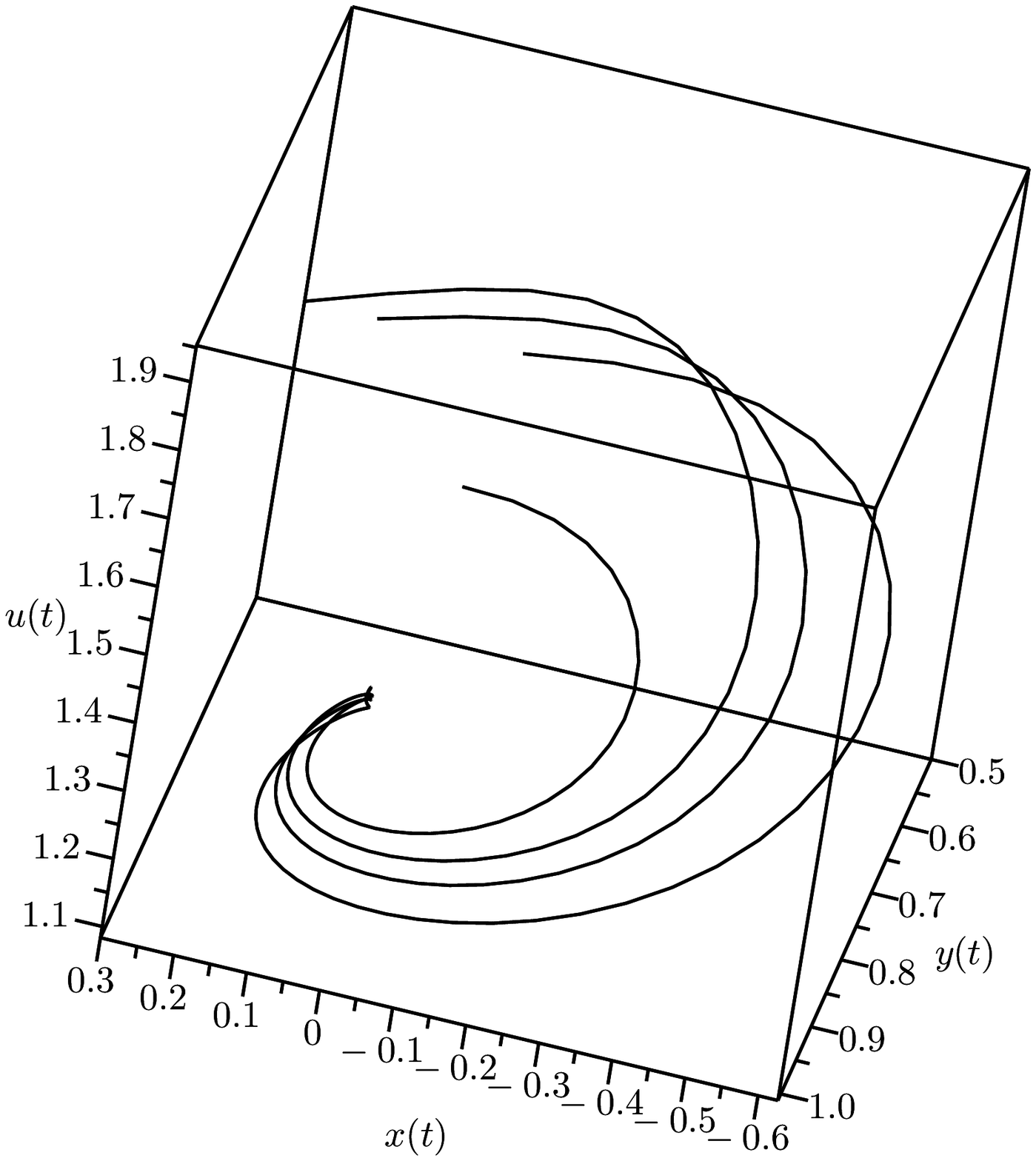} \vspace{2.5cm}\includegraphics{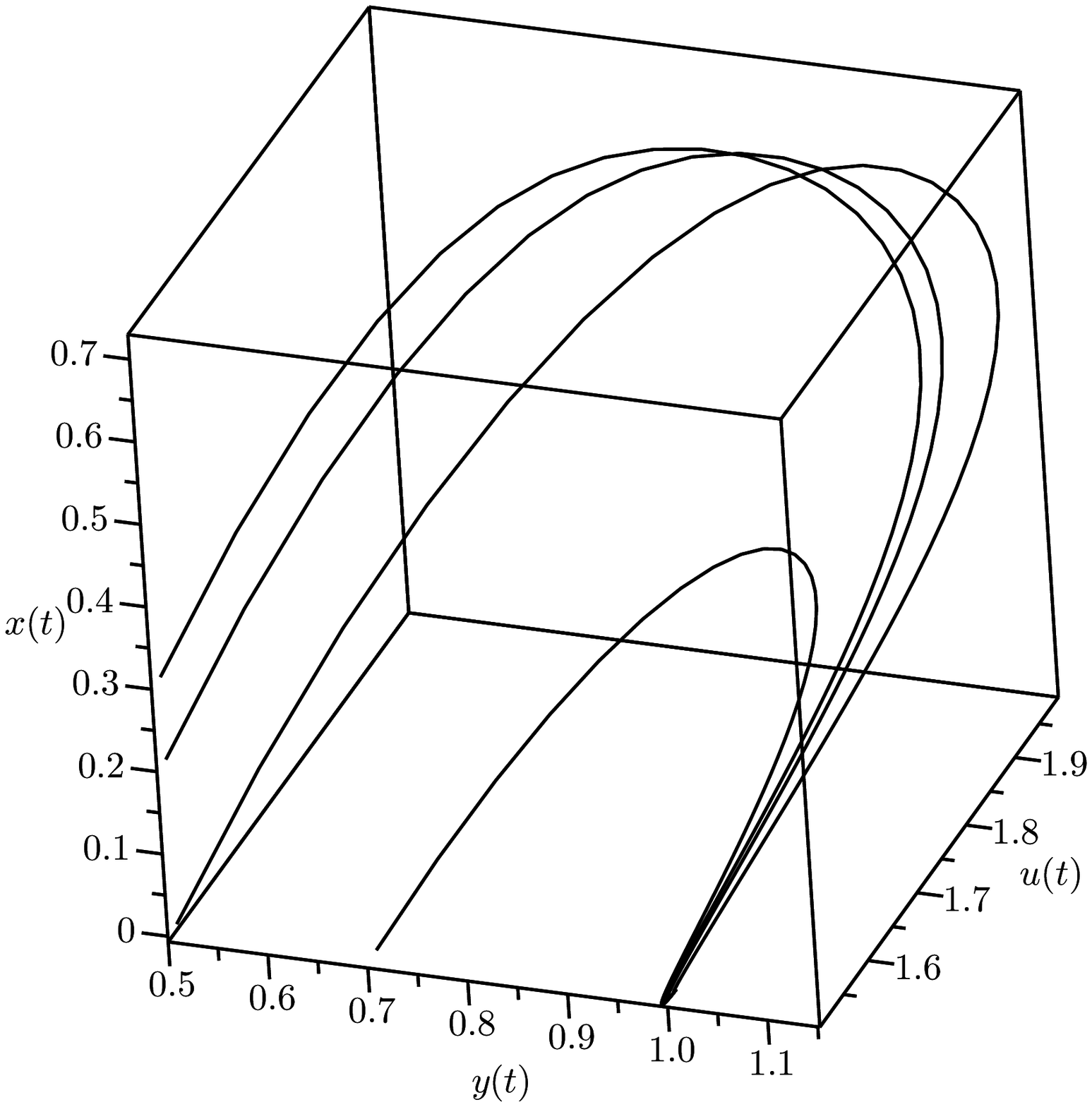}\vspace{2.5cm}

\caption{\small {3-dimensional phase-space trajectories of the
cosmological scenario (16)-(18) with stable attractor $P_{3}$ for
$\lambda=2$, $\alpha=1$ and $\eta=-\frac{1}{2}$ (left) and
$\lambda=-2$, $\alpha=0.5$ and $\eta=\frac{1}{2}$ (right). }}
\end{center}
\end{figure}

\begin{figure}[htp]
\begin{center}
\includegraphics{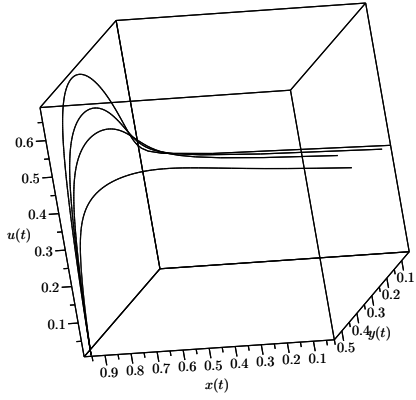} \vspace{2.5cm}

\caption{\small {3-dimensional phase-space trajectories of the
cosmological scenario (16)-(18) with stable attractor $P_{4}$ for
$\lambda=3$, $\alpha=-2$ and $\eta=\frac{1}{2}$.}}
\end{center}
\end{figure}

\begin{figure}[htp]
\begin{center}
\includegraphics{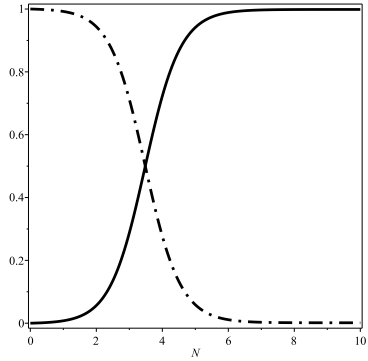} \vspace{6cm}

\caption{\small {Evolution of $\Omega_{\phi}$ (solid) and
$\Omega_{m}$(dotdashed) with $\lambda=0.3$, $\alpha=-1.5$ and
$\eta=\frac{1}{2}$. The initial conditions are $x_{i}=10^{-8}$,
$y_{i}=3.7 \times 10^{-2}$ and $u_{i}=10^{-7}$. The corresponding
values of $\Omega_{\phi}$ and $\Omega_{m}$ at the present epoch
($N\simeq 4$) are $\Omega_{\phi}\approx0.72$ and
$\Omega_{m}\approx0.28$.}}
\end{center}
\end{figure}

\section{Appendix: Elements of perturbation matrix $\Sigma$ and its eigenvalues at critical points}
The components of matrix $\Sigma$ which represents the coefficients
of the linearized perturbation equations are given by
\begin{multline}
\Sigma_{11}=-2x_{c}\left[(3-s_{c})\eta\,\mu_{c}^{-1}\,u_{c}^{2}y_{c}^{-2}+\sqrt{3}\left(\lambda\,y_{c}-\sqrt{3}\,x_{c}\right)\right]-3\mu_{c}^{-3}\eta
u_{c}^{2}y_{c}^{-2}\left(2+\eta^{2}\mu_{c}^{-3}u_{c}^{4}y_{c}^{-2}\right)^{-1}\\
\left[s_{c}\mu_{c}^{-1}\eta^{2}x_{c}
u_{c}^{4}y_{c}^{-2}-(\gamma-1)\left(\mu_{c}^{3}x_{c}y_{c}^{2}-\eta
u_{c}^{2}\right)+\mu_{c} x_{c}y_{c}^{2} +\eta
u_{c}^{2}\left(3x_{c}^{2}-1+\frac{2\sqrt{3}}{3}x_{c}y_{c}(\alpha-\lambda)-3\mu_{c}^{-1}\eta
x_{c}u_{c}^{2}y_{c}^{-2}\right)\right]\\ -\mu_{c}^{-1}\eta x_{c}
u_{c}^{2}y_{c}^{-2}(3-s_{c})-3\mu_{c}^{-2},
\end{multline}

\begin{multline}
 \Sigma_{12}=\sqrt{3}\lambda\left(1-x_{c}^{2}\right)-2\eta
 \mu_{c}^{-3}u_{c}^{2}y_{c}^{-3}-\eta
 \mu_{c}^{-3}u_{c}^{2}y_{c}^{-2}\\
 \left(2+\eta^{2}\mu_{c}^{-3}u_{c}^{4}y_{c}^{-2}\right)^{-1}
 \left[2\eta^{2}\mu_{c}^{-3}s_{c} u_{c}^{4}y_{c}^{-3}+6\mu_{c} y_{c}\left(x_{c}^{2}-\gamma\right)+\eta\,u_{c}^{2}
 \left(\sqrt{3}\left(x_{c}^{2}(\alpha-\lambda)+\lambda\right)+6\eta\,\mu_{c}^{-3}
 u_{c}^{2}y_{c}^{-3}\right)\right],
\end{multline}

\begin{multline}
 \Sigma_{13}=2(3-s_{c})\eta\mu_{c}^{-3}u_{c} y_{c}^{-2}-2\eta\mu_{c}^{-3}u_{c}^{2} y_{c}^{-2}
 \left(2+\eta^{2}\mu_{c}^{-3}u_{c}^{4}y_{c}^{-2}\right)^{-1}\\
\left[3\eta\,u_{c}\left((\gamma-1)x_{c}-\mu_{c}^{-2}x_{c}+\frac{\sqrt{3}}{3}y_{c}\left(x_{c}^{2}(\alpha-\lambda)+\lambda\right)+2\eta\,\mu_{c}^{-3}
 u_{c}^{2}y_{c}^{-2}\right)-2\eta^{2}\,\mu_{c}^{-3}s_{c}
 u_{c}^{3}y_{c}^{-2}\right],
\end{multline}

\begin{multline}
 \Sigma_{21}=-\frac{\sqrt{3}}{2}\lambda y_{c}^{2}-3y_{c}\left(2+\eta^{2}\mu_{c}^{-3}u_{c}^{4}y_{c}^{-2}\right)^{-1}\\
 \left[s_{c}\mu_{c}^{-1}\eta^{2}x_{c}
u_{c}^{4}y_{c}^{-2}-(\gamma-1)\left(\mu_{c}^{3}x_{c}y_{c}^{2}-\eta
u_{c}^{2}\right)+\mu_{c} x_{c}y_{c}^{2}+\eta
u_{c}^{2}\left(3x_{c}^{2}-1+\frac{2\sqrt{3}}{3} x_{c}
y_{c}(\alpha-\lambda)-3\mu_{c}^{-1}\eta
x_{c}u_{c}^{2}y_{c}^{-2}\right)\right],\\
\end{multline}

 \begin{multline}
\Sigma_{22}= s_{c}-\sqrt{3}\lambda
x_{c} y_{c}+y_{c}\left(2+\eta^{2}\mu_{c}^{-3}u_{c}^{4}y_{c}^{-2}\right)^{-1}\\
 \left[2\eta^{2}\mu_{c}^{-3}s_{c} u_{c}^{4}y_{c}^{-3}+6\mu_{c} y_{c}\left(x_{c}^{2}-\gamma\right)+\eta\,u_{c}^{2}
 \left(\sqrt{3}\left(x_{c}^{2}(\alpha-\lambda)+\lambda\right)+6\eta\,\mu_{c}^{-3}
 u_{c}^{2}y_{c}^{-3}\right)\right],
 \end{multline}

\begin{multline}
 \Sigma_{23}=2 y_{c} \left(2+\eta^{2}\mu_{c}^{-3}u_{c}^{4}y_{c}^{-2}\right)^{-1}\\
\left[3\eta\,u_{c}\left((\gamma-1)x_{c}-\mu_{c}^{-2}x_{c}+\frac{\sqrt{3}}{3}y_{c}\left(x_{c}^{2}(\alpha-\lambda)+\lambda\right)+2\eta\,\mu_{c}^{-3}
 u_{c}^{2}y_{c}^{-2}\right)-2\eta^{2}\,\mu_{c}^{-3}s_{c}
 u_{c}^{3}y_{c}^{-2}\right],
\end{multline}

 \begin{multline}
 \Sigma_{31}=\frac{\sqrt{3}}{2}\alpha y_{c}u_{c}-\frac{3}{2} u_{c}\left(2+\eta^{2}\mu_{c}^{-3}u_{c}^{4}y_{c}^{-2}\right)^{-1}\\
 \left[s_{c}\mu_{c}^{-1}\eta^{2}x_{c}
u_{c}^{4}y_{c}^{-2}-(\gamma-1)\left(\mu_{c}^{3}x_{c}y_{c}^{2}-\eta
u_{c}^{2}\right)+\mu_{c} x_{c}y_{c}^{2}+\eta u_{c}^{2}
\left(3x_{c}^{2}-1+\frac{2\sqrt{3}}{3} x_{c}
y_{c}(\alpha-\lambda)-3\mu_{c}^{-1}\eta
x_{c}u_{c}^{2}y_{c}^{-2}\right)\right],\\
\end{multline}

\begin{multline}
\Sigma_{32}=\frac{\sqrt{3}}{2}\alpha
x_{c} u_{c}+\frac{1}{2} u_{c}\left(2+\eta^{2}\mu_{c}^{-3}u_{c}^{4}y_{c}^{-2}\right)^{-1}\\
 \left[2\eta^{2}\mu_{c}^{-3}s_{c} u_{c}^{4}y_{c}^{-3}+6\mu_{c} y_{c}\left(x_{c}^{2}-\gamma\right)+\eta\,u_{c}^{2}
 \left(\sqrt{3}\left(x_{c}^{2}(\alpha-\lambda)+\lambda\right)+6\eta\,\mu_{c}^{-3}
 u_{c}^{2}y_{c}^{-3}\right)\right],
 \end{multline}

 \begin{multline}
 \Sigma_{33}=\frac{1}{2}\left(\sqrt{3}\,\alpha\,x_{c}\,y_{c}+s_{c}\right)+\frac{1}{2}u_{c} \left(2+\eta^{2}\mu_{c}^{-3}u_{c}^{4}y_{c}^{-2}\right)^{-1}\\
\left[3\eta\,u_{c}\left((\gamma-1)x_{c}-\mu_{c}^{-2}x_{c}+\frac{\sqrt{3}}{3}y_{c}\left(x_{c}^{2}(\alpha-\lambda)+\lambda\right)+2\eta\,\mu_{c}^{-3}
 u_{c}^{2}y_{c}^{-2}\right)-2\eta^{2}\,\mu_{c}^{-3}s_{c}
 u_{c}^{3}y_{c}^{-2}\right].
\end{multline}
In the above expressions by $s_{c}$ and $\mu_{c}$ we mean evaluation
of them at critical points i.e
\begin{equation}
\mu_{c}=1/\sqrt{1-x_{c}^{2}},
\end{equation}
and
\begin{multline}
s_{c}=3\,\left(2+\eta^{2}\mu_{c}^{-3}u_{c}^{4}y_{c}^{-2}\right)^{-1}\\
\left[\gamma-(\gamma-1)\left(
 \mu_{c}\,y_{c}^{2}-\eta\,u_{c}^{2}\,x_{c}\right)-\mu_{c}^{-1}\,y_{c}^{2}+\eta\,u_{c}^{2}\left(-\mu_{c}^{-2}\,x_{c}+\frac{\sqrt{3}}{3}y_{c}
 \left(x_{c}^{2}(\alpha-\lambda)+\lambda\right)+\eta\,\mu_{c}^{-3}u_{c}^{2}y_{c}^{-2}\right)\right].
\end{multline}
Although elements of $\Sigma$ seem complicated, inserting the
explicit critical points shown in Table 1, the matrix $\Sigma$
acquires a simple form that allows for calculation of its
eigenvalues. The eigenvalues of the matrix $\Sigma$ (namely
$\nu_{1}$, $\nu_{2}$, $\nu_{3}$), for each critical point are as
follows:\\
points $P_{1}$ and $P_{2}$:
\begin{equation}
 \nu_{1}=\frac{3\,\gamma\left(\lambda+2\alpha\right)}{4\,\lambda},
  \:\:\:\:\:\:\: \nu_{2,3}=\frac{3\,\left(\gamma\lambda-2\lambda\pm\sqrt{17\gamma^{2}\lambda^{2}-20\gamma\lambda^{2}+4\lambda^{2}+
  48\gamma^{2}\sqrt{1-\gamma}}\right)}{4\,\lambda}
\end{equation}
Point $P_{3}$:
\begin{equation}
 \nu_{1,2}=\frac{-3\,\left(\lambda^{2}+6\pm\sqrt{-3\lambda^{4}-12\lambda^{2}+36-48\lambda\alpha-8\alpha\lambda^{3}}\right)}
 {2\,\left(6+\lambda^{2}\right)},\:\:\:\:\:\:\: \nu_{3}=-3\,\gamma.
 \label{34}
\end{equation}

\end{document}